\documentclass[submission,copyright,creativecommons]{eptcs}
\providecommand{\event}{iWIGP2011} 
\usepackage{breakurl}              

\newcommand{\red}[1]{#1}

\title{Synthesizing Systems with Optimal Average-Case Behavior for Ratio Objectives}

\author{Christian von Essen
\institute{VERIMAG\\ Grenoble, France}
\institute{EDMSTII\\Université Joseph Fourier, Grenoble, France}
\email{christian.vonessen@imag.fr}
\and
Barbara Jobstmann
\institute{CNRS/VERIMAG\\ Grenoble, France}
\email{barbara.jobstmann@imag.fr}
}
\def\titlerunning{Synthesizing Systems with Optimal Average-Case Behavior for Ratio Objectives}

\def\authorrunning{C. von Essen \& B. Jobstmann}

\usepackage{amsmath}
\usepackage{amsthm}
\usepackage{amsfonts}
\usepackage{algorithm2e}
\usepackage{tikz}
\usetikzlibrary{automata}
\usepackage{subfigure}

\newtheorem{defi}{Definition}
\newtheorem{lem}{Lemma}
\newtheorem{thm}{Theorem}

\begin{document}

\renewcommand{\implies}{\Rightarrow}

\newcommand{\R}{\ensuremath{\mathbb{R}}}
\newcommand{\N}{\ensuremath{\mathbb{N}}}
\newcommand{\B}{\ensuremath{\mathbb{B}}}
\newcommand{\true}{1}
\newcommand{\false}{0}

\newcommand{\seq}{\word}
\newcommand{\run}{\ensuremath{\rho}}
\newcommand{\Runs}{\ensuremath{\Omega}}
\newcommand{\finRun}{\ensuremath{v}}
\newcommand{\infRun}{\ensuremath{\rho}}
\newcommand{\cone}[1]{\ensuremath{\gamma}(#1)}

\newcommand{\distribs}{\ensuremath{\mathcal{D}}}
\newcommand{\pOutcomes}{\ensuremath{\Omega}}
\newcommand{\pAlgebra}{\ensuremath{\mathcal{F}}}
\newcommand{\pMeasure}{\ensuremath{\mu}}
\newcommand{\pSpace}{\ensuremath{\mathcal{P}}}
\newcommand{\prob}{\ensuremath{p}}

\newcommand{\nVisits}{\ensuremath{\nu}}
\newcommand{\Prob}{\ensuremath{P}}

\newcommand{\MDP}{\ensuremath{\mathcal{M}}}
\newcommand{\mdpStates}{\ensuremath{\Sigma}}
\newcommand{\mdpState}{\ensuremath{\sigma}}
\newcommand{\mdpStart}{\ensuremath{\mdpState_0}}
\newcommand{\mdpProb}{\ensuremath{p}}
\newcommand{\strategy}{\ensuremath{\pi}}

\newcommand{\mStates}{\ensuremath{S}}
\newcommand{\mState}{\ensuremath{s}}
\newcommand{\mStart}{\ensuremath{\mState_0}}
\newcommand{\mProb}{\ensuremath{p}}
\newcommand{\LabelsOf}{\lambda}
\newcommand{\ActionsOf}{\tilde{A}}

\newcommand{\bSystem}{\ensuremath{\mathcal{B}}}
\newcommand{\aSystem}{\ensuremath{\mathcal{A}}}
\newcommand{\aStates}{\ensuremath{Q}}
\newcommand{\aSafeStates}{\ensuremath{F}}
\newcommand{\aStart}{\ensuremath{q_0}}
\newcommand{\aState}{\ensuremath{q}}
\newcommand{\aTrans}{\ensuremath{\delta}}
\newcommand{\aOut}{\ensuremath{\Lambda}}
\newcommand{\cost}{\ensuremath{c}}

\newcommand{\eModel}{f_e}

\newcommand{\Actions}{\ensuremath{A}}
\newcommand{\Labels}{\ensuremath{L}}
\newcommand{\action}{\ensuremath{a}}
\newcommand{\Label}{\ensuremath{l}}
\newcommand{\payoff}{\ensuremath{{\cal R}}}
\newcommand{\startAction}{\ensuremath{A_0}}

\newcommand{\lang}[1]{{\cal L}_{#1}}
\newcommand{\alphabet}{\Sigma}
\newcommand{\word}{w}
\newcommand{\emptyword}{\epsilon}
\newcommand{\quant}{\psi}
\newcommand{\quali}{\varphi}

\newcommand{\transducer}[1]{{\cal O}_{#1}}
\newcommand{\system}{\ensuremath{{\cal S}}}
\newcommand{\sStates}{\ensuremath{S}}
\newcommand{\sState}{\ensuremath{s}}
\newcommand{\sStart}{\ensuremath{s_0}}
\newcommand{\sTrans}{\ensuremath{\delta}}
\newcommand{\sOut}{\ensuremath{\tau}}
\newcommand{\eValue}[2]{\ensuremath{\mathbb{E}#1[#2]}}
\newcommand{\V}[3]{\mbox{Value}_{#2}^{#1}(#3)}

\newcommand{\rDef}[1]{Def.~\ref{#1}}
\newcommand{\rLem}[1]{Lem.~\ref{#1}}
\newcommand{\rEqn}[1]{Eqn.~\ref{#1}}
\newcommand{\rFig}[1]{Figure~\ref{#1}}
\newcommand{\rSec}[1]{Section~\ref{#1}}

\newcommand{\bjnote}[1]{{\red{\bf BJ: #1}}}
\newcommand{\venote}[1]{{\bf CVE: #1}}
\newcommand{\rev}[1]{{\bf REVIEWER: #1}}

\newcommand{\distribution}{\mu}

\newcommand{\set}[1]{\{#1\}}

\maketitle

\begin{abstract}
We show how to automatically construct a system that satisfies a given
logical specification and has an optimal average behavior with respect
to a specification with ratio costs.

When synthesizing a system from a logical specification, it is often
the case that several different systems satisfy the specification.  In
this case, it is usually not easy for the user to state formally which
system she prefers.  Prior work proposed to rank the correct systems
by adding a quantitative aspect to the specification.  A desired
preference relation can be expressed with (i) a quantitative language,
which is a function assigning a value to every possible behavior of a
system, and (ii) an environment model defining the desired
optimization criteria of the system, e.g., worst-case or average-case
optimal.

In this paper, we show how to synthesize a system that is optimal for
(i) a quantitative language given by an automaton with a ratio
cost function, and (ii) an environment model given by a labeled Markov
decision process.  The objective of the system is to minimize the
expected (ratio) costs.  The solution is based on a reduction to
Markov Decision Processes with ratio cost functions
which do not require that the costs in the denominator are strictly
positive.  We find an optimal strategy for these using a fractional
linear program.
\end{abstract}

\section{Introduction}

Quantitative analysis techniques are usually used to measure
quantitative properties of systems, such as timing, performance, or
reliability (cf.~\cite{cacm,prism,uppaal}).  We use quantitative
reasoning in the classically Boolean contexts of verification and
synthesis because they allow us to distinguish systems with respect to
``soft constraints'' like robustness~\cite{BloemGHJ09} or default
behavior~\cite{BloemCHJ09}.  This is particularly helpful in
synthesis, where a system is automatically derived from a
specification, because quantitative specifications allow us to guide
the synthesis tool towards a desired implementation.

In this paper we show how quantitative specifications based on
ratio objectives can be used to guide the synthesis process.  In
particular, we present a technique to synthesize a system with an
average-case behavior that satisfies a logical specification and
optimizes a quantitative objective given by a ratio objective.

The synthesis problem can be seen as a game between two players: the
system and the environment (the context in which the system operates).
The system has a fixed set of interface variables with a finite domain
to interact with its environment.  The variables are partitioned into
a set of input and output variables.  The environment can modify the
set of input variables.  For instance, an input variable can indicate
the arrival of some packet on a router on a given port or the request
of a client to use a shared resource.  Each assignment to the input
variables is a possible move of the environment in the synthesis game.
The system reacts to the behavior of the environment by changing the
value of the output variables.  An assignment to the output variables
is called an action of the system and describes a possible move of the
system in the synthesis game.  E.g., the system can grant a shared
resource to Client~$C$ by setting a corresponding output variable.
Environment and system change their variables in turns.  In every
step, first the system makes modification to the output variables,
then the environment changes the input variables.  The sequence of
variable evaluations built up by this interplay is evaluated with
respect to a specification.  A logical (or qualitative) specification
maps every sequence to $\true$ or $\false$, indicating whether the
sequence satisfies the specification or not.  For example, a sequence
of evaluations in which the system grants a shared resource to two
clients at the same time is mapped to $\false$ if the specification
requires mutual exclusive access to this resource.  The aim of the
system in the synthesis game is to satisfy the specification
independent of the choices of the environment.
There might be several systems that can achieve this goal for a given
specification. Therefore, Bloem et al.~\cite{BloemCHJ09} proposed to
add a quantitative specification in order to rank the correct systems.
A quantitative specification maps every infinite sequence of variable
evaluations to a value indicating how desirable this behavior is.
\red{In this paper, we study quantitative specifications resulting
  from ratio objectives.  The idea is that a behavior of the system is
  mapped to two infinite sequences of values. The first sequence
  refers to events that were ``good'' for the system, while the second
  sequence refers to ``bad'' events within a behavior. For instance,
  consider a server processing requests from several clients. If the
  server receives a request it can be seen as a bad event, since it
  requires the server to process the request. On the other hand,
  every handled request is clearly a good
  event. Intuitively, the ratio objectives computes the long-run ratio
  between the sum of bad and the sum of good events.  This ratio is
  the value of a behavior.}
A system can be seen as a set of behaviors.  \red{We can assign a
  value to a system by taking, e.g., the worst or the average value
  over all its behaviors. Given a way to evaluate a system,} we can
ask for a system that optimizes this value, i.e., a system that
achieves a better value than any other system.  Taking the worst value
over the possible behaviors corresponds to assuming that the system is
in an adversary environment.  The average value is computed with
respect to a probabilistic model of the
environment~\cite{ChatterjeeHJS10}.  In the average-case synthesis
game, the environment player is replaced by a probabilistic player
that is playing according to the probabilistic environment model.

In this paper, we present the first average-case synthesis algorithm
for specifications that evaluate a behavior of the system with respect
to the ratio of two cost functions~\cite{BloemCHJ09}.  This ratio
objective allows us, e.g., to ask for a system that optimizes the
ratio between requests and acknowledgments in a server-client system.
For the average-case analysis, we present a new environment model,
which is based on Markov decision processes and generalizes the one
in~\cite{ChatterjeeHJS10}.  We solve the average-case synthesis
problem with ratio objective by reduction to Markov decision processes
with ratio cost functions. For unichain Markov Decision Processes with
ratio cost functions, we present a solution based on linear
programming.

\paragraph{Related Work.}
Researchers have considered a number of formalisms for quantitative
specifications~\cite{Alur09,CCHKM05,CAFHMS06,CDH08,dA98,DiscountingTheFuture,DrosteGastin07,DrosteKR08,KL07}
but most of them (except for~\cite{BloemGHJ09}) do not consider
long-run ratio objectives.  In~\cite{BloemGHJ09}, the environment is
assumed to be adversary, while we assume a probabilistic environment
model.  Regarding the environment model, there have been several
notions of metrics for probabilistic systems and games proposed in the
literature~\cite{dAMRS07,DGJP99}.  The metrics measure the distance of
two systems with respect to all temporal properties expressible in a
logic, whereas we (like~\cite{ChatterjeeHJS10}) uses the quantitative
specification to compare systems wrt the property of interest.  In
contrast to~\cite{ChatterjeeHJS10}, we use ratio objectives and a more
general environment model.  Our environment model is the same as the
one used for control and synthesis in the presence of uncertainty
(cf.~\cite{Baier,CourcoubetisY90,BdA95}). However, in this context
usually only qualitative specifications are considered.  MDPs with
long-run average objectives are well studied.  The
books~\cite{FV96,puterman} present a detailed analysis of this topic.
Cyrus Derman~\cite{derman62} studied MDPs with a fractional objective.
This work differs in two aspects from ours: first, Derman requires
that the payoff of the cost function of the denominator is always
strictly positive and second, the objective function used
in~\cite{derman62} is already given in terms of the expected cost of
the first cost function to the expected cost of the second cost
functions and not in terms of a single trace.
\red{De Alfaro~\cite{deAlfaro97} studies a model that is similar to
  ours but does not consider the synthesis problem.}
Finally, we would like to note that the two choices we have in a
quantitative synthesis problem, namely the choice of the quantitative
language and the choice of environment model are the same two choices
that appear in weighted automata and max-plus algebras
(cf.~\cite{Droste09,Gaubert97,Cuning79}).

\section{Preliminaries}

\paragraph{Words, Qualitative and Quantitative Languages.}

Given a finite alphabet $\alphabet$, a \emph{word}
$\word=\word_0\word_1\dots$ is a finite or infinite sequence of
elements of $\alphabet$.  We use $\word_i$ to denote the $(i+1)$-th
element in the sequence.  If~$\word$ is finite, then $|\word|$ denotes
the length of $\word$, otherwise $|\word|$ is infinity.  We denote the
empty word by $\emptyword$, i.e., $|\emptyword|=0$.  We use
$\alphabet^*$ and $\alphabet^\omega$ to denote the set of finite and
infinite words, respectively.  Given a finite word
$\word\in\alphabet^*$ and a finite or infinite word $v\in \alphabet^*
\cup \alphabet^\omega$, we write $\word v$ for the concatenation of
$\word$ and $v$.  A \emph{qualitative language} $\quali$ is a function
$\quali:\alphabet^\omega \to \B$ mapping every infinite word to 1
or 0.  Intuitively, a qualitative language partitions the set of
words into a set of good and a set of bad traces.  A
\emph{quantitative language}~\cite{CDH08} $\quant$ is a function
$\quant:\alphabet^\omega\to\R^+ \cup \set{\infty}$ associating to each
infinite word a value from the extended non-negative reals.

\paragraph{Specifications and automata with cost functions.}

An \emph{automaton} is a tuple $\aSystem=(\alphabet,
\aStates,\aStart,\aTrans,\aSafeStates)$, where $\alphabet$ is a finite
\emph{alphabet}, $Q$ is a finite set of \emph{states}, $\aStart\in
\aStates$ is an \emph{initial state}, $\delta:Q \times \alphabet \to
\aStates$ is the \emph{transition function}, and
$\aSafeStates\subseteq\aStates$ is a set of \emph{safe states}.
We
use $\sTrans^*: \sStates \times \Labels^* \to \sStates$ to denote
the closure of $\sTrans$ over finite words. Formally, given a word
$\word=\word_0\dots\word_n\in \alphabet^*$, $\sTrans^*$ is defined
inductively as $\sTrans^*(q,\epsilon)=q$, and
$\sTrans^*(q,w)=\sTrans(\sTrans^*(q,\word_0\dots\word_{n-1}),\word_n)$.
We use $|\aSystem|$ to denote the size of the automaton.

The \emph{run} $\run$ of $\aSystem$ on an infinite word
$\word=\word_0\word_1\word_2\dots\in \alphabet^\omega$ is an infinite
sequence of states $\aState_0\aState_1\aState_2\dots$ such that
$\aState_0$ is the initial state of $\aSystem$ and $\forall i\ge 0:
\delta(\aState_i,w_i)=\aState_{i+1}$ holds. The run $\run$ is called
\emph{accepting} if for all $i\ge 0$, $\aState_i\in\aSafeStates$.  A
word $\word$ is accepting if the corresponding run is accepting. The
\emph{language of~$\aSystem$}, denoted by $\lang{\aSystem}$, is the
qualitative language $\lang{\aSystem}:\alphabet^\omega \to \B$
mapping all accepting words to $\true$ and non-accepting words to
$\false$, i.e., $\lang{\aSystem}$ is the characteristic function of
the set of all accepting words of ${\aSystem}$.
  We
assume without loss of generality that $\aStates\setminus\aSafeStates$
is closed under~$\aTrans$, i.e., $\forall s \in
\aStates\setminus\aSafeStates, \forall a \in \alphabet: \aTrans(s,a)
\in \aStates\setminus\aSafeStates$.  \red{Note that every automaton
  can be modified to meet this assumption by (i) adding a new
  state~$\aState_{\bot}$ with a self-loop for every letter and (ii)
  redirecting every transition starting from $\aStates \setminus
  \aSafeStates$ to the new state~$\aState_{\bot}$.}
The modified automaton accepts the same language as the original
automaton.

Given an automaton
$\aSystem=(\alphabet,\aStates,\aStart,\aTrans,\aSafeStates)$, a
\emph{cost function} $\cost: \aStates \times \alphabet\to\N$ is a
function that maps every transition in $\aSystem$ to a non-negative
integer.  We use automata with cost functions and \emph{objective
  functions} to define quantitative languages (or properties).
Intuitively, the objective function tells us how to summarize the
costs along a run.  
Given an automation $\aSystem$ and two cost functions
$\cost_1,\cost_2$, the \emph{ratio objective}~\cite{BloemGHJ09}
computes the ratio between the costs seen along a run of $\aSystem$ on
a word $\word=\word_0\word_1\word_2\dots\in \alphabet^\omega$:
\begin{equation}
  \label{eqn:payoff}
  \payoff(\word) := \lim_{m \to \infty} \liminf_{l \to \infty}
  \frac{\sum_{i = m}^{l}
    \cost_1(\delta^*(\aStart,\word_0\dots\word_i),\word_{i+1})}{1 +
    \sum_{i = m}^{l}
    \cost_2(\delta^*(\aStart,\word_0\dots\word_i),\word_{i+1})}
\end{equation}

\noindent
The ratio objective is a generalization of the long-run average
objective (also known as mean-payoff objective, cf.~\cite{Zwick96}).
We use $\payoff^{\aSystem}_{\frac{\cost_1}{\cost_2}}$ to denote the
quantitative language defined by $\aSystem$, $\cost_1$, $\cost_2$, and
the ratio objective function. If $\aSystem$, $\cost_1$, or $\cost_2$
are clear from the context, we drop them.

Intuitively, $\payoff$ computes the long-run ratio between the costs
accumulated along a run.  The first limit allows us to ignore a finite
prefix of the run, which ensures that we only consider the long-run
behavior.  The $1$ in the denominator avoids division by $0$, if the
accumulated costs are $0$ and has no effect if the accumulated costs
are infinite.
We need the limit inferior here because the sequence of the limit
might not converge. 
\red{Consider the sequence $\run=q^1 r^2 q^4r^8q^{16}\dots$, where
  $q^k$ means that the State~$q$ is visited $k$-times.  Assume
  State~$q$ and State~$r$ have the following costs: $\cost_1(q) = 0$,
  $\cost_2(q) = 1$, $\cost_1(r) = 1$ and $\cost_2(r) = 1$.  Then, the
  value of~$\run_0\dots \run_i$ will alternate between $0$ and $1$
  with increasing~$i$ and hence the sequence for $i\to\infty$ will not
  converge.} The limit inferior of this sequence is $0$.

\paragraph{Finite-state system and Correctness}

A \emph{finite-state system} $\system=(\sStates, \Labels, \sStart,
\Actions, \sTrans, \sOut)$ consists of the automaton
$\aSystem=(\Labels,
\sStates,\sStart, \sTrans, \sStates)$\footnote{Note that the last
  element of this tuple is the set of safe states, i.e., every
state is safe.}, an \emph{output (or action)
  alphabet} $\Actions$, and an \emph{output function} $\sOut: \sStates
\to \Actions$ assigning to each state of the system a letter from the
output alphabet. The alphabet of the automaton $\Labels$ is called the
\emph{input alphabet} of the system.  Given an input word $w$, the
\emph{run of the system $\system$ on the word $w$} is simply the run
of $\aSystem$ on the word $w$.  For every word $\word$ over the input
alphabet, the system produces a word over the joint input/output
alphabet.  We use $\transducer{\system}$ to denote the function
mapping input words to the joint input/output word, i.e, given an
input word $\word=\word_0\word_1\dots\in
\Labels^\omega$, $\transducer{\system}(\word)$ is the sequence of
tuples $(\Label_0,\action_0)(\Label_1,\action_1)\dots\in
(\Labels\times\Actions)^\omega$ such that (i) $\Label_i=\word_i$ for
all $i\ge0$, (ii) $\action_0=\sOut(\sStart)$, and (iii) for all $i>0$,
$\action_i=\sOut(\sTrans^*(\sStart,\word_0\dots\word_{i-1})))$ holds.

Given a system $\system$ with input alphabet $\Labels$ and output
alphabet $\Actions$, and an automaton $\aSystem$ with alphabet
$\alphabet=\Labels\times\Actions$, we say that the system $\system$
\emph{satisfies the specification} $\aSystem$, denoted
$\system\models\aSystem$, if for all input words, the joint
input/output word produced by the system $\system$ is accepted by the
automaton $\aSystem$, i.e.,
$\forall
\word\in\Labels^\omega: (\lang{\aSystem}\circ
\transducer{\system})(\word)=1,
$ 
where $\circ$ denotes the function composition operator.

\paragraph{Probability space.}

We use the standard definitions of probability spaces.
A \emph{probability space} is given by a tuple $\pSpace :=
(\pOutcomes, \pAlgebra, \pMeasure)$, where $\pOutcomes$ is the set of
\emph{outcomes or samples}, $\pAlgebra \subseteq 2^{\pOutcomes}$ is
the $\sigma$-algebra defining the set of \emph{measurable events}, and
$\pMeasure \in \pAlgebra \to [0, 1]$ is a \emph{probability measure}
assigning a probability to each event such that $\pMeasure(\pOutcomes)
= 1$ and for each countable set $E_1, E_2, \dots \in \pAlgebra$ of
disjoint events we have $\pMeasure(\bigcup E_i) = \sum \pMeasure(E_i)$.  Recall
that, since $\pAlgebra$ is a $\sigma$-algebra, it satisfies the
following three conditions: (i) $\emptyset \in \pAlgebra$, (ii) $E \in
\pAlgebra$ implies $\pOutcomes \setminus E \in \pAlgebra$ for any
event $E$, and (iii) the union of any countable set of events
$E_1, E_2, \dots \in \pAlgebra$ is also in $\pAlgebra$,
i.e., $\bigcup E_i \in \pAlgebra$.
Given a measurable function $f: \pAlgebra \to \R \cup \{ +\infty,
-\infty \}$, we use $\eValue{_\pSpace}{f}$ to
denote the expected value of $f$ under $\pMeasure$, i.e.,
\begin{equation}
    \eValue{_\pSpace}{f} =
    \int_{\pOutcomes} f~d\pMeasure
\end{equation}
If $\pSpace$ is clear from the context we drop the subscript or
replace it with the structure that defines $\pSpace$.  The integral
used here is the Lebesgue Integral, which is commonly used to define
the expected value of a random variable.  Note that the expected value
is always defined if the function $f$ maps only to values in $\R^+
\cup \{\infty \}$.

\paragraph{Markov chains and Markov decision processes (MDP).}

Let $\distribs(S) := \{ \prob: S \to [0, 1] \mid \sum_{s \in S}
\prob(s) = 1 \}$ be the \emph{set of probability distributions} over a
set $S$.

A \emph{Markov decision process} is a tuple $\MDP = (\mStates,
\mStart, \Actions, \ActionsOf, \mProb)$, where $\mStates$ is a finite
set of \emph{states}, $\mStart\in \mStates$ is an \emph{initial
  state}, $\Actions$ is the finite set of \emph{actions},
$\ActionsOf:\mStates \to 2^{\Actions}$ is the \emph{enabled action
  function} defining for each state $\mState$ the set of enabled
actions in $\mState$, and $\mProb: \mStates \times \Actions \to
\distribs(\mStates)$ is the probabilistic \emph{transition function}.
For technical convenience we assume that every state has at least one
enabled action, i.e., $\forall \mState\in\mStates:
|\ActionsOf(\mState)| \ge 1$.  If $|\ActionsOf(\mState)|=1$ for all
states $\mState \in \mStates$, then $\MDP$ is called a \emph{Markov
  chain (MC)}.  In this case, we omit $\Actions$ and $\ActionsOf$ from
the definition of $\MDP$.  Given a Markov chain $\MDP$, we say that
$\MDP$ is \emph{irreducible} if every state can be reached from any
other. We say that it is \emph{unichain} if it has at most one
maximal set of states that can reach it other. We call an MDP
unichain if every strategy induces a unichain MC.

An \emph{$\Labels$-labeled Markov decision process} is a tuple $\MDP =
(\mStates, \mStart, \Actions, \ActionsOf, \mProb, \LabelsOf)$, where
$(\mStates, \mStart, \Actions, \ActionsOf, \mProb)$ is a Markov
decision process and $\LabelsOf:\mStates\to\Labels$ is a labeling
function such that $\MDP$ is deterministic with respect to
$\LabelsOf$, i.e, for all states $\mState, \mState', \mState''$ and
every action $\action$ such
that $\mState' \not = \mState''$,  $\mProb(\mState,\action)(\mState') > 0$ and
$\mProb(\mState,\action)(\mState'') > 0$ we have  $\LabelsOf(\mState')\ne
\LabelsOf(\mState'')$.  Since we use $\Labels$-labeled Markov decision
process to represent the behavior of the environment, we require that
in every state all actions are enabled, i.e., $\forall
\mState\in\mStates: \ActionsOf(s)=\Actions$.

\paragraph{Sample runs and strategies}
A \emph{(sample) run} $\infRun$ of $\MDP$ is an infinite sequence of
tuples $(\mState_0,\action_0)(\mState_1,\action_1)\dots\in
(\mStates\times\Actions)^\omega$ of states and actions such that for
all $i \ge 0$, (i)~$\action_i\in\ActionsOf(\mState_i)$ and
(ii)~$\mProb(\mState_i,\action_i)(\mState_{i+1})>0$.  We use $\Runs$
to denote the set of all runs, and $\Runs_{\mState}$ for the set of
runs starting at state $\mState$.
A \emph{finite run} of $\MDP$ is a prefix of some infinite run.  To
avoid confusion, we use $\finRun$ to refer to a finite run. Given a
finite run $\finRun$, the set $\cone{\finRun}:=\{ \infRun \in\Runs
\mid \exists \infRun' \in \Runs: \infRun=\finRun\infRun'\}$ of all
possible infinite extensions of $\finRun$ is called the \emph{cone
  set} of $\finRun$.  We use the usual extension of $\cone{\cdot}$ to
sets of finite words.

A \emph{strategy} is a function $\strategy:(\mStates
\times\Actions)^*\mStates \to \distribs(\Actions)$ that assigns a
probability distribution to all finite sequences in $(\mStates
\times\Actions)^{*}\mStates$.  A strategy must refer only to enabled
actions, i.e., for all sequences $\seq \in (\mStates \times \Actions)^*$,
states $\mState\in\mStates$, and actions $\action\in\Actions$, if
$\strategy(\seq\mState)(\action)>0$, then action $\action$ has to be
enabled in $\mState$, i.e., $\action\in\ActionsOf(\mState)$.
A strategy $\strategy$ is \emph{pure} if for all finite sequences
$\seq\in (\mStates \times \Actions)^*$ and for all states
$\mState\in\mStates$, there is an action $\action\in\Actions$ such
that $\strategy(\seq\mState)(\action)=1$.  A \emph{memoryless}
strategy is independent of the history of the run, i.e., for all
$\seq,\seq'\in(\mStates \times \Actions)^*$ and for all
$\mState\in\mStates$, $\strategy(\seq\mState)=\strategy(\seq'\mState)$
holds. A memoryless strategy can be represented as function
$\strategy:\mStates\to\distribs(\Actions)$. A pure and memoryless
function can be represented by a function
$\strategy:\mStates\to\Actions$ mapping states to actions.
An MDP $\MDP = (\mStates, \mStart, \Actions, \ActionsOf, \mdpProb)$
together with a pure and memoryless strategy
$\strategy:\mStates\to\Actions$ defines the Markov chain
$\MDP^{\strategy} = (\mStates, \mStart, \Actions,
\ActionsOf_{\strategy}, \mdpProb)$, in which only the actions
prescribed in the strategy $\strategy$ are enabled, i.e.,
$\ActionsOf_{\strategy}(\mState) = \{ \strategy(\mState) \}$.  Note
that every finite-state system~$\system$ with input
alphabet~$\mStates$ and output alphabet~$\Actions$ \red{that refers
  only to enabled actions} can be viewed as a strategy for~$\MDP$.
Vice-versa, \red{an MDP with} a pure and memoryless
strategy~$\strategy$ defines a finite state
system~$\system_{\strategy}^{\MDP}$ with input alphabet~$\mStates$ and
output alphabet~$\Actions$.

\paragraph{Induced probability space, objective function, and optimal
  strategies.}
An MDP $\MDP = (\mStates, \mStart, \Actions, \ActionsOf, \mProb)$
together with a strategy $\strategy$ and a state $\mState\in\mStates$
induces a probability space $\pSpace_{\MDP,\mState}^{\strategy} =
(\pOutcomes_{\MDP,\mState}^{\strategy},
\pAlgebra_{\MDP,\mState}^{\strategy},
\pMeasure_{\MDP,\mState}^{\strategy})$ over the cone sets of the
runs starting in $\mState$.  Hence, $\pOutcomes_{\MDP,\mState}^{\strategy} =
\mStates^{\omega}$.  The probability measure of a cone set is the
probability that the MDP starts from state $\mState$ and follows the
common prefix under the strategy $\strategy$.  By convention
$\pSpace_{\MDP}^{\strategy} := \pSpace_{\MDP,\mStart}^{\strategy}$. If
$\MDP$ is a Markov chain, then $\strategy$ is fixed (since there is
only one available action in every state), and we simply write
$\pSpace_{\MDP}$.

An \emph{objective function} of $\MDP$ is a measurable function
$f:(\mStates\times \Actions)^\omega\to\R^+ \cup \set{\infty}$ that
maps runs of~$\MDP$ to values in $\R^+ \cup \set{\infty}$.  We use
$\eValue{_{\MDP, \mState}^{\strategy}}{f}$ to denote the expected
value of $f$ wrt the probability space induced by the MDP $\MDP$, a
strategy $\strategy$, and a state $\mState$.

We are interested in a strategy that has the least expected value for
a given state.  Given an MDP~$\MDP$ and a state $\mState$, a strategy
$\strategy$ is called \emph{optimal} for objective $f$ and state
$\mState$ if
$\eValue{_{\MDP,\mState}^{\strategy}}{f}=\min_{\strategy'}\;\eValue{_{\MDP,
  \mState}^{\strategy'}}{\payoff},
$ 
where~$\strategy'$ ranges over all possible strategies.

Given an MDP $\MDP= (\mStates, \mStart, \Actions, \ActionsOf, \mProb)$
and two cost function $\cost_1:\mStates\times\Actions\to\N$ and
$\cost_2:\mStates\times\Actions\to\N$, the \emph{ratio
  payoff value} is the function $\payoff:(\mStates\times \Actions)^\omega\to\R^+ \cup
\set{\infty}$ mapping every run $\run$ to a value in $\R^+ \cup
\set{\infty}$ as follows:
\begin{equation}
  \payoff_{\frac{\cost_1}{\cost_2}}(\run) := \lim_{m \to \infty} \liminf_{l \to \infty}
  \frac{\sum_{i = m}^{l} \cost_1(\run_i)}{1 + \sum_{i = m}^{l} \cost_2(\run_i)}
\end{equation}
We drop the subscript $\frac{\cost_1}{\cost_2}$ if $\cost_1$ and
$\cost_2$ are clear from the context.

\section{Synthesis with Ratio Objective in Probabilistic Environments}
\label{sec:synthesis}
In this section, we first present a variant of the quantitative
synthesis problem introduced in~\cite{BloemCHJ09}.  Then, we show how
to solve the synthesis problem with safety and ratio
specifications in a probabilistic environment described by an MDP.

The quantitative synthesis problem with probabilistic environments
asks to construct a finite-state system $\system$ that satisfies a
qualitative specification and optimizes a quantitative specification
under the given environment.
The specifications are qualitative and quantitative languages over
letters in $(\Labels\times\Actions)$, where $\Labels$ and $\Actions$
are the input and output alphabet of $\system$, respectively.

In order to compute the average behavior of a system, we \red{assume} a model
of the environment.  In \cite{ChatterjeeHJS10}, the environment model
is a probability space $\pSpace=(\Labels^\omega,\pAlgebra,\pMeasure)$
over the input words $\Labels^\omega$ of the system defined by a
finite $\Labels$-labeled Markov chain.  This model assumes that the
behavior of the environment is independent of the behavior of the
system, which restricts the modeling possibilities.  
\red{For instance, a client-server system, in which a client increases
  the probability of sending a request if it has not been served in
  the previous step, cannot be modeled using this approach.}
Therefore, \emph{our environment model} is a function
$\eModel$ that maps every system $f_s:\Labels^*\to\Actions$ to a
probability space $\pSpace=(\Labels^\omega,\pAlgebra,\pMeasure)$ over
the input words $\Labels^\omega$.  Note that every finite-state system
defines such a system function $f_s$ but not vice versa.  To describe
a particular environment model $\eModel$, we use a finite
$\Labels$-labeled Markov decision process. Once we have an environment
model, we can define what it means for a system to satisfy a
specification under a given environment.

\begin{defi}[Satisfaction]
\label{def:satisfaction}
Given a finite-state system $\system$ with alphabets~$\Labels$ and
$\Actions$, a qualitative specification $\quali$ over alphabet
$\Labels\times\Actions$, and an environment model $\eModel$, we say
that \emph{$\system$ satisfies $\quali$ under $\eModel$} (written
$\system\models_{\eModel} \quali$\footnote{Note that
  $\system\models_{\eModel} \quali$ and $\system\models \quali$
  coincide if (i) $\quali$ is prefix-closed (which is the case for the
  specifications, we consider here), and (ii) $\eModel(\system)$
  assigns, for every finite word $\word\in\Labels^*$, a positive
  probability to the set of infinite words $\word\Labels^\omega$.})
iff $\system$ satisfies $\quali$ with probability~$1$, i.e.,
  \[
  \eValue{_{\eModel(\system)}}{\quali\circ\transducer{\system}}=1.
  \]
\end{defi}

Recall that $\transducer{\system}$ denotes the function that maps
\red{input words to  joint input/output words}, and that $\quali$ is a
qualitative specification,
\red{which maps (input/output) words to~$0$ or~$1$.}
Hence,
$\quali\circ\transducer{\system}$ denotes 
\red{the} function that maps
\red{an input sequence to~$1$ if the behavior of the system $\system$
  for this input word satisfies the specification~$\quali$. Otherwise,
  the input word is mapped to~$0$.} 
The function $\eValue{_{\eModel(\system)}}{f}$ of some measurable
function~$f$ denotes the expected value of $f$ under the probability
distribution induced by the system $\system$ under the environment
model $\eModel$.  \red{Hence, Definition~\ref{def:satisfaction} says
  that a system satisfies a specification under a probabilistic
  environment model if almost all behaviors of the system satisfy the
  specification, i.e., the probability that the system misbehaves
  is~$0$.}

Next, we define the value of a system with respect to a specification
under an environment model and what it means for a system to optimize
a specification. Then, we are ready to define the quantitative
synthesis problem.

\begin{defi}[Value of a system]
\label{def:value}
  Given a finite-state system $\system$ with alphabets~$\Labels$ and
  $\Actions$, a qualitative~($\quali$) and a quantitative
  specification ($\quant$) over alphabet $\Labels\times\Actions$,
  and an environment model $\eModel$, the \emph{value of $\system$ with
    respect to $\quali$ and $\quant$ under $\eModel$} is defined as the
  expected value of the function $\quant \circ \transducer{\system}$
  in the probability space $\eModel(\system)$, if $\system$ satisfies
  $\quali$, and~$\infty$ otherwise.  Formally,
\[
  \V{\eModel}{\quali\quant}{\system} :=
  \begin{cases}
    \eValue{_{\eModel(\system)}}{\quant \circ \transducer{\system}} &
    \mbox{if } \system\models_{\eModel}\quali,\\
    \infty & \mbox{otherwise.}\\
  \end{cases}
\]
If $\quali$ is the set of all words, then we write
$\V{\eModel}{\quant}{\system}$. Furthermore, we say \emph{$\system$ optimizes
  $\quant$ wrt $\eModel$}, if $\V{\eModel}{\quant}{\system} \le
\V{\eModel}{\quant}{\system'}$ for all systems~$\system'$.
\end{defi}

\begin{defi}[Quantitative realizability and synthesis problem]
  Given a qualitative specification $\quali$ and a quantitative
  specification $\quant$ over the alphabets~$\Labels\times\Actions$
  and an environment model $\eModel$, the \emph{realizability problem} asks
  to decide if there exists a finite-state system $\system$ with
  alphabets~$\Labels$ and $\Actions$ such that
  $\V{\eModel}{\quali\quant}{\system} \ne \infty$.
  The \emph{synthesis problem} asks to construct a finite-state system
  $\system$ (if it exists) s.\ t.
  \begin{enumerate}
  \item $\V{\eModel}{\quali\quant}{\system} \ne \infty$ and
  \item $\system$ optimizes $\quant$ wrt $\eModel$.
  \end{enumerate}
\end{defi}

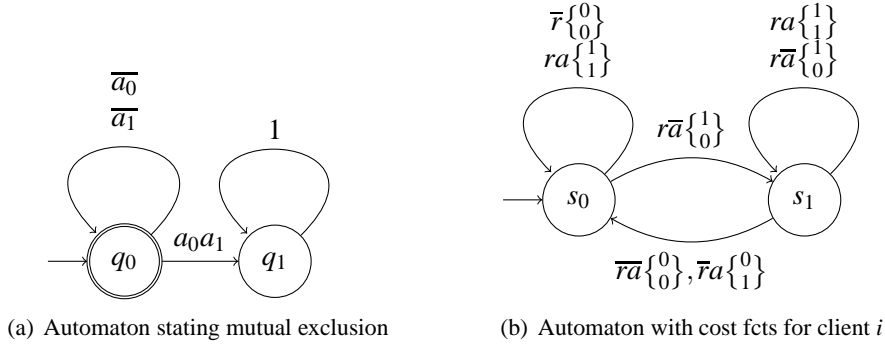
\begin{figure}[bt]
  \centering
\subfigure[Automaton stating mutual exclusion]{
  \begin{tikzpicture}[node distance=2cm,auto]
    \node[state, initial,accepting,initial text=] (q_0)                {$q_0$};
    \node[state]                    (q_1) [right of=q_0] {$q_1$};

    \path[->] (q_0) edge [loop,above] node {
      $\begin{array}{c}
        \overline{a_0}\\
        \overline{a_1}
        \end{array}$} (q_0)
                    edge              node {$a_0a_1$}                     (q_1);
    \path[->] (q_1) edge [loop,above] node {$\true$}                              (q_1);

 \end{tikzpicture}
\label{fig:mutual}
}
\subfigure[Automaton with cost fcts for client~$i$]{
  \begin{tikzpicture}[node distance=3cm,auto]
    \node[state, initial,initial text=] (s_0) {$s_0$};
    \node[state]          (s_1) [right of=s_0] {$s_1$};

    \path[->] (s_0) edge [loop, above] node {$\begin{array}{c}
                                                 \overline{r}             { 0 \brace 0 }\\
                                                 r           a            { 1 \brace 1 }
                                              \end{array}$} (s_0)
                    edge [bend left]   node {$r \overline{a} { 1 \brace 0 }$} (s_1);
    \path[->] (s_1) edge [loop,above]  node {$\begin{array}{c}
                                                  ra { 1 \brace 1 }\\
                                                  r \overline{a} { 1 \brace 0 }
                                              \end{array}$ } (s_1)
                    edge [bend left]   node {$
                                               \overline{r}\overline{a} { 0 \brace 0 },
                                               \overline{r} a           { 0 \brace 1 }
                                              $} (s_0);

  \end{tikzpicture}


\label{fig:req-ack-ratio}}
\caption{Specifications for the client-server example}
\end{figure}


In the following, we give an example of a quantitative synthesis
problem. 
\paragraph{Server-client example.}
Consider a server-client system with two clients and one server.  Each
server-client interface consists of two variables $r_i$ (request) and
$a_i$ (acknowledge). Client~$i$ sends a request by setting $r_i$ to
$\true$. The server acknowledges the request by setting $a_i$ to
$\true$. We require that the server does not acknowledge both clients
at the same time.  Hence, our qualitative specification demands mutual
exclusion. \rFig{fig:mutual} shows an automaton stating the mutual
exclusion property for $a_1$ and $a_2$. Edges are labeled with
\red{sets of} evaluations of $a_1$ and $a_2$, 
\red{e.g., $\overline{a}_1$ states that $a_1$ has to be~$\false$ and
  $a_2$ can have either value, $\true$ and $\false$.}
States drawn with a double circle are safe states.  Among all systems
satisfying the mutual exclusion property, we ask for a system that
minimizes the average ratio between requests and useful
acknowledgments.  An acknowledge is useful if it is sent as a response
to a request.  To express this property, we can give a quantitative
language defined by an automaton with two cost functions
($\cost_1,\cost_2$) and the ratio objective (\rEqn{eqn:payoff}).
\rFig{fig:req-ack-ratio} shows an automaton labeled with tuples
representing the two cost functions $\cost_1$ and $\cost_2$ for one
client.  The first component of the tuples represents cost
function~$\cost_1$, the second component defines cost
function~$\cost_2$.  The cost function $\cost_1$ is $1$, whenever we
see a request.  The cost function $\cost_2$ is $1$, when we see a
``useful'' acknowledge, which is an acknowledge that matches an
unacknowledged request. E.g., every acknowledge in state $s_1$ is
useful, since the last request has not been acknowledged yet. In state
$s_0$ only acknowledgments that answer a direct request are useful and
get cost~$1$ (in the second component). This corresponds to a server
with a buffer that can hold exactly one request and that gets outdated
after two steps and has to be dropped.
State $s_1$ says that there is a request in the
buffer. If there is no acknowledgment while the
machine is in this state, then the request is lost. This means that a
request has to be acknowledged in the step it is received or in the
step after that.

\begin{figure}[bt]
\centering
\subfigure[MDP of one client]{
 \begin{tikzpicture}[node distance=3cm,auto]
   \node[state,initial,initial text=]  (r_0) {$\overline{r}$};
   \node[state]  (r_1) [right of=r_0] {$r$};
   \node[state]  (r_0') [right of=r_1] {$\overline{r}$};

   \path[->] (r_0) edge  [loop below]       node { $a,\overline{a}; 1/2$ }(r_0)
                   edge                     node { $a,\overline{a}; 1/2$ }(r_1);
   \path[->] (r_1) edge  [loop below]       node { $\begin{array}{c}
                                                       \overline{a}; 3/4\\
                                                                 a;  1/2
                                                    \end{array}$ }       (r_1)
                   edge  [bend left]        node { $a; 1/2$ }            (r_0)
                   edge  [bend right,below] node { $\overline{a}; 1/4$ } (r_0');
   \path[->] (r_0')edge  [loop below]       node { $\overline{a}; 1/8$ } (r_0')
                   edge  [bend right]       node { $\begin{array}{c}
                                                        a;            1/2 \\
                                                        \overline{a}; 7/8
                                                    \end{array}$ }
                                                  (r_1);
   \draw [->] (r_0') .. controls (5.5,1.3) and (0.5,1.3) .. (r_0);
    \node at (3,1.3) {$a;            1/2$};
\end{tikzpicture}
 \label{fig:mdp}
}
\subfigure[Implementation of a server for two clients.  State labeling:
  {$\sOut(m_0)=a_1 \overline{a}_2$ and $\sOut(m_1)=\overline{a}_1
    a_2$}]{
    \begin{tikzpicture}[node distance=3cm,auto]
    \node[state, initial,initial text=] (s_0) {$m_0$}; 
    \node[state]          (s_1) [right of=s_0] {$m_1$}; 

    \path[->] (s_0) edge [loop, above] node {$\begin{array}{c}
        r_1\\
        \overline{r}_2
        \end{array}$} (s_0)
                    edge [bend left]   node {$\overline{r}_1 r_2$} (s_1);
    \path[->] (s_1) edge [loop, above] node {$\overline{r}_1 r_2$} (s_1)
                    edge [bend left]   node {$\begin{array}{c}
                        r_1\\
                        \overline{r}_2
                      \end{array}$} (s_0);
  \end{tikzpicture}
  \label{fig:system}
  }
  \caption{Specifications and implementation for the client-server example}
\end{figure}
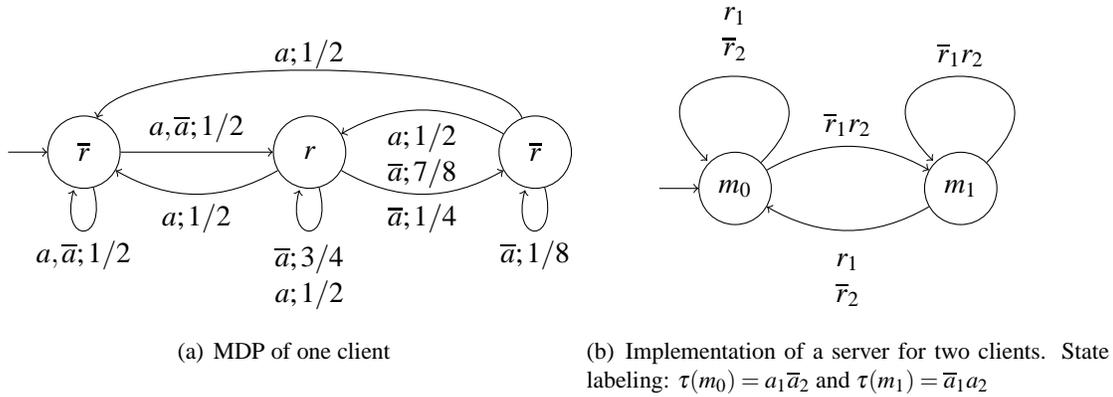

Assume we know the expected behavior of the clients. E.g., in every
step, Client~$1$ is expected to send a request with probability $0.5$
independent of the acknowledgments.  Client~$2$ changes its behavior
based on the acknowledgments.  We can describe the behavior of
Client~$2$ by the labeled MDP shown in~\rFig{fig:mdp}. In the
beginning the chance of getting a request from this client is $0.5$.
Once it has sent a request, i.e., it is in state $r$, the probability
of sending a request again is very high until at least one
acknowledgment is given. This is modeled by action $\overline{g}$ at
state $r$ having a probability of $3/4$ to get into state $r$ again,
and a probability of $1/4$ to not send a request in the next step. In
this case, we move to the right $\overline{r}$ state. In this state,
the probability of receiving a request from this client in the next
step is even $7/8$. This means that if this client does not receive
an acknowledgment after having sent a request, then the possibility of
receiving another request from this client in the next two steps is $1
- 1/4 * 1/8 = 31/32$.


Consider the finite-state system $\system$ shown in \rFig{fig:system}.
It is an implementation of a server for two clients. The system has
two states $m_0$ and $m_1$ labeled with $a_1 \overline{a}_2$ and
$\overline{a}_1 a_2$, respectively.  We can compute the value of
$\system$ using the following two lemmas
(\rLem{lm:safety-reduction}, \rLem{lm:ratio-reduction}).

\newpage

\begin{lem}
\label{lm:safety-reduction}
Given (i) a finite-state system $\system$ with alphabets~$\Labels$
and $\Actions$, (ii) an automaton $\aSystem$ with alphabet
$\Labels\times\Actions$, and (iii) an $\Labels$-labeled MDP $\MDP$
defining an environment model for $\system$, there exists a Markov
chain $\MDP_c$ and two cost functions $\cost_1$ and $\cost_2$ such
that
\[
\system\models_{\MDP}\lang{\aSystem}
\quad\mathop{\iff}^{\scriptsize\mbox{\rDef{def:satisfaction}}}\quad
\eValue{_{\MDP}^{\system}}{\lang{\aSystem}\circ\transducer{\system}}=1
\quad\iff\quad
 \eValue{_{\MDP_c}}{\payoff_{\frac{\cost_1}{\cost_2}}}=0
\]
\end{lem}

\emph{Proof idea:} The Markov chain $\MDP_c$ is constructed by taking
the synchronous product of $\system$, $\aSystem$, and~$\MDP$.  In
every state $(s,q,m)\in (\sStates_{\system} \times \aStates_{\aSystem}
\times \mStates_{\MDP})$, we take the action $a\in\Actions$ given by
the labeling function of the system $\sOut(s)$ and move to a successor
state for every input label $l\in\Labels$ such that there exists a
state $m'$ in the MDP $\MDP$ with $\LabelsOf(m')=l$ and
$p(m,a)(m')>0$. The corresponding successor states of the system- and
the automaton-state components are $s'=\sTrans_{\system}(s,l)$ and
$q'=\aTrans_{\aSystem}(q,(l,a))$.  The probability distribution of
$\MDP_c$ is taken from the $\MDP$-component.  The two cost functions
are defined as follows: for state $(s,q,m)$ and an action $a$ we set 
$\cost_1((s,q,m),a)=0$ and $\cost_2((s,q,m),a)=1$, if $q$ is a safe
state in $\aSystem$, otherwise $\cost_1((s,q,m),a)=1$ and
$\cost_2((s,q,m),a)=0$. Intuitively, since the non-safe states of
$\aSystem$ are (by definition) closed under $\aTrans_{\aSystem}$ and
all actions in this set have the same cost, they all have the same
value, namely $\infty$, so does every state from which there is a
positive probability to reach this set.\footnote{Note that instead of
  an MDP with ratio objective, we could have also set up a two-player
  safety game here.}

\begin{lem}
  Given (i) a finite-state system $\system$ with alphabets~$\Labels$
  and $\Actions$, (ii) an automaton $\aSystem$ with alphabet
  $\Labels\times\Actions$ with two cost functions $\cost_1$ and
  $\cost_2$, and (iii) a $\Labels$-labeled MDP $\MDP$ defining an
  environment model for $\system$, there exists a Markov chain
  $\MDP_c$ and two cost functions $d_1$ and $d_2$ such that
\[
\V{\MDP}{\payoff_{\frac{\cost_1}{\cost_2}}}{\system}
\quad\mathop{=}^{\scriptsize\mbox{\rDef{def:value}}}\quad
\eValue{_{\MDP}^{\system}}{\payoff_{\frac{\cost_1}{\cost_2}}\circ\transducer{\system}}
\quad=\quad\eValue{_{\MDP_c}}{\payoff_{\frac{d_1}{d_2}}}
\]
\label{lm:ratio-reduction}
\end{lem}

\emph{Proof idea:} The construction is the same as the one for
\rLem{lm:safety-reduction} except for the cost functions. The cost
functions are simply copied from the component referring to the
automaton, e.g., given a state $(s,q,m)\in (\sStates_{\system} \times
\aStates_{\aSystem} \times \mStates_{\MDP})$ and an action
$a\in\Actions$, $d_1((s,q,m),a)=\cost_1(q)$ and
$d_2((s,q,m),a)=\cost_2(q)$.

In Section~\ref{sec:mdp-ratio}, we show how to compute an optimal
value for MDPs with ratio objectives in polynomial time. Since Markov
chains with ratio objectives are a special case of MDPs with ratio
objectives, we can first use \rLem{lm:safety-reduction} to check if
$\system\models_{\MDP}\lang{\aSystem}$. If the check succeeds, we then
use \rLem{lm:ratio-reduction} to compute the value
$\V{\MDP}{\payoff_{\frac{\cost_1}{\cost_2}}}{\system}$.  This
algorithm leads to the following theorem.

\begin{thm}[System value]
  Given 
  a finite-state system $\system$ with alphabets~$\Labels$ and
  $\Actions$, 
  an automaton $\aSystem$ with alphabet $\Labels\times\Actions$
  defining a qualitative language, 
  an automaton $\bSystem$ with alphabet $\Labels\times\Actions$ and
  two cost functions $\cost_1$ and $\cost_2$ defining a quantitative
  language, and
  a $\Labels$-labeled MDP $\MDP$ defining an environment model,
  we can compute value of $\system$ with respect to $\lang{\aSystem}$
  and $\payoff_{\frac{\cost_1}{\cost_2}}$ under
  $\pSpace_{\MDP}^{\system}$ in time polynomial in the maximum of
  $|\system| \cdot |\aSystem| \cdot |\MDP|$ and 
  $|\system|\cdot |\bSystem| \cdot 
  |\MDP|$.  
\end{thm}

\red{In order to synthesize an optimal system, we construct an MDP
  from the environment model, the quantitative, and qualitative
  specifications similar to the constructions in
  \rLem{lm:safety-reduction} and~\ref{lm:ratio-reduction}.  Any
  optimal strategy for this MDP with a value different from~$\infty$
  corresponds to a system that satisfies the qualitative specification
  and optimizes the quantitative specifications.  In the next section,
  we will show that MDPs with ratio objectives have pure memoryless
  optimal strategies. Therefore, we need to consider only such
  strategies that are pure and memoryless.}
\red{Given a pure and memoryless strategy, we build the corresponding
  system as follows:} we reduce the set of enabled actions in each
state to the single action specified by the strategy.  \red{In each
  state, the enabled action defines the output function of the
  system.}
Instead of deciding the next state probabilistically, 
\red{the system moves from one to the next state depending on the
  chosen input value.}

\red{In the next section we show how to compute an optimal strategy
  for a given MDP in time polynomial in the number of states.  This
  result together with construction above leads to the following
  theorem.}

\begin{thm}[Synthesis]
  Given
  an automaton $\aSystem$ with alphabet $\Labels\times\Actions$
  defining a qualitative language ,
  an automaton $\bSystem$ with alphabet $\Labels\times\Actions$ and
  two cost functions $\cost_1$ and $\cost_2$ defining a quantitative
  language, and
  a $\Labels$-labeled MDP $\MDP$ defining an environment model,
  we can compute an optimal system $\system$ with respect to $\lang{\aSystem}$
  and $\payoff_{\frac{\cost_1}{\cost_2}}$ in time polynomial in
  $|\aSystem| \cdot |\bSystem| \cdot |\MDP|$.
\end{thm}

\section{Calculating the best strategy}
\label{sec:mdp-ratio}

In this section we will first outline a proof showing that for every
MDP there is a pure and memoryless optimal strategy for our payoff function.
To this end, we argue how the proof given by \cite{gimbert} can be
adapted to our case. After that we will show how we can calculate
an optimal pure and memoryless strategy.

\subsection{Pure and memoryless strategies suffice}

In~\cite{gimbert}, Gimbert proved that in an MDP any payoff function
mapping to $\R$ that is submixing and prefix independent admits optimal
pure and memoryless strategies.  Since our payoff function $\payoff$ may also
take the value~$\infty$, we cannot apply the result immediately.
However, since $\payoff$ maps only to non-negative values and the set
of measurable functions is closed under addition, multiplication,
limit inferior and superior and division, provided that the divisor is
not equal to $0$, the expected value of $\payoff$ is always defined
and the theory presented in~\cite{gimbert} also applies in this case.
Furthermore, to adapt the proof of~\cite{gimbert} to minimizing the
payoff function instead of maximizing it, one only needs to inverse
the used inequalities and replace $\max$ by $\min$. What remains to
show is that $\payoff$ fulfills the following two properties.

\begin{lem}[$\payoff$ is submixing and prefix independent]
  Let $\MDP = (\mStates, \Actions, \ActionsOf, \mProb)$ be a
  MDP and $\run$ be a run.
  \begin{enumerate}
  \item For every $i \ge 0$ the prefix of $\run$ up to $i$
    does not matter, i.e., $\payoff(\run) = \payoff(\run_i \run_{i+1} \dots)$.

  \item For every sequence of non-empty words $u_0, v_0, u_1, v_1
    \dots \in (\Actions \times \mStates)^+$ such that $\run = u_0 v_0
    u_1 v_1 \dots$ we have that the payoff of the sequence is greater
    than or equal to the minimal payoff of sequences $u_0 u_1 \dots$
    and $v_0 v_1 \dots$, i.e., $\payoff(\run) \ge \min\{ \payoff(u_0
    u_1 \dots), \payoff(v_0 v_1 \dots) \}$.
  \end{enumerate}

  \begin{proof}
    The first property follows immediately from the first limit in the
    definition of $\payoff$.

    For the second property we partition $\N$ into $U$ and $V$ such
    that $U$ contains the indexes of the parts of $\run$ that belong
    to a $u_k$ for some $k \in \N$ and such that $V$ contains the
    other indexes. Formally, we define $U := \bigcup_{i \in \N} U_i$ where $U_0
    := \{ k \in \N \mid 0 \le k < |u_0| \}$ and $U_i := \{
    max(U_{i-1}) + |v_{i-1}| + k \mid 1 \le k \le |u_i| \}$. Let $V :=
    U \setminus \N$ be the other indexes.

    Now we look at the payoff from $m$ to $l$ for some $m \le l \in
    \N$, i.e. $\payoff_m^l := (\sum_{i = m \dots l} \cost_1(\run_i))/(1 +
      \sum_{i = m \dots l} \cost_2(\run_i))$. We can divide the sums into two
    parts, the one belonging to $U$ and the one belonging to $V$ and
    we get
    \[
    \payoff_m^l = \frac{\displaystyle
      \phantom{1 +}
      \left( \sum_{i \in \{m \dots l\} \cap U} \cost_1(\run_i) \right) +
      \left( \sum_{i \in \{m \dots l\} \cap V} \cost_1(\run_i) \right)}{\displaystyle
      1 +
      \left( \sum_{i \in \{m \dots l\} \cap U} \cost_2(\run_i) \right) +
      \left( \sum_{i \in \{m \dots l\}\cap V} \cost_2(\run_i) \right)}
    \]
    We now define the sub-sums between the parentheses as $u_1 := \sum_{i
      \in \{m \dots l\} \cap U} \cost_1(\run_i)$, $u_2 := \sum_{i \in \{m \dots
      l\} \cap U} \cost_2(\run_i)$, $v_1 := \sum_{i \in \{m \dots l\} \cap V}
    \cost_1(\run_i)$ and $v_2 := \sum_{i \in \{m \dots l\} \cap V} \cost_2(\run_i)$. Then
    we receive
    \[
    \payoff_m^l = \frac{u_1 + v_1}{1 + u_2 + v_2}
    \]

    \noindent We will now show \[
    \payoff_m^l \ge \min \left\{ \frac{u_1}{u_2 + 1}, \frac{v_1}{v_2 +
      1} \right\}
    \]
    Without loss of generality we can assume
    $u_1/(u_2 + 1) \ge v_1/(v_2 + 1)$, then we have to show that
    \[
    \frac{u_1 + v_1}{1 + u_2 + v_2} \ge \frac{v_1}{v_2 + 1}.
    \]
    This holds if and only if $(u_1 + v_1)(1 + v_2) = u_1 + v_1 +
    u_1v_2 + v_1v_2 \ge v_1(1 + u_2 + v_2) = v_1 + v_1u_2 + v_1v_2$
    holds. By subtracting $v_1$ and $v_1v_2$ from both sides we
    receive $u_1 + u_1v_2 = u_1(1 + v_2) \ge u_2v_1$. If $u_2$ is
    equal to 0 then this holds because $u_1$ and $v_2$ are greater
    than or equal to 0. Otherwise, this holds if and only
    if $u_1/u_2 \ge v_1/(1 + v_2)$ holds. In general, we have
    $u_1/u_2 \ge u_1/(u_2 + 1)$. 
    From the assumption we have $u_1/(u_2 + 1) \ge v_1/(v_2 + 1)$
    and hence $u_1/u_2 \ge v_1/(v_2 + 1)$. The original
    claim follows because we have shown this for any pair of $m$
    and $l$.
  \end{proof}
\end{lem}
    
\begin{thm}[There is always a pure and memoryless optimal strategy]
  For each MDP with the ratio payoff function, there
  is a pure and memoryless optimal strategy.

  \begin{proof}
    See \cite{gimbert}
  \end{proof}
\end{thm}

\subsection{Reduction of MDP to a Linear Fractional Program}

In this section, we show how to calculate a pure and memoryless optimal
strategy for an MDP with ratio objective by reducing the problem to a
fractional linear programming problem.  A fractional linear
programming problem is similar to a linear programming problem, but
the function that one wants to optimize is the fraction of two linear
functions.  A fractional linear programming problem can be reduced to
a series of conventional linear programming problems to calculate
the optimal value.

We present the reduction only for unichain MDPs.  The extension to
general MDPs is based on end-components \cite{deAlfaro97} and the fact that
end-components have an optimal unichain strategy.

Our reduction uses the fact that an MDP with a pure and memoryless strategy
induces a Markov chain and that the runs of a Markov chain have a
special property akin to the law of large numbers, which we can use to
calculate the expected value.

\begin{defi}[Random variables of MCs]
  Let $\prob^n(\mState)$ be the probability of being in state
  $\mState$ at step $n$ and let $\prob^*(\mState) := lim_{n \to
    \infty} \frac{1}{n} \sum_{i = 0}^{n-1} \prob^i(\mState)$. This is
  called the \emph{Cesaro limit} of $\prob^n$. Let further
  $\nVisits_{\mState}^n$ denote the \emph{number of visits} to state
  $\mState$ up to time $n$.
\end{defi}

We have the following lemma describing the long-run behavior of
unichain Markov chains
\cite{AFirstCourseInStochasticModels,MarkovChains}.

\begin{lem}[Expected number of visits of a state and well-behaved runs]
  \label{lem:ENVisits}
  For every infinite run of a unichain Markov chain, the fraction of visits to a
  specific state $\mState$ equals $\prob^*(\mState)$ almost surely, i.e.,
  $\Prob(\lim_{l \to \infty} \frac{\nVisits_{\mState}^l}{l} =
  \prob^*(\mState)) = 1$. We call the set of runs that have this
  property \emph{well-behaved}.
\end{lem}

When we calculate the expected payoff, we only need to consider
well-behaved words as shown in the following lemma.

\begin{lem}
  Let $N$ denote the set of runs that are not well-behaved. Then
  \[
  \eValue{_\MDP}{\payoff} = \int_{\pOutcomes_{\MDP} \setminus N} \payoff ~ d\pMeasure_{\MDP}
  \]

  \begin{proof}
    The probability measure of the set of well-behaved words is
    1. Hence the probability measure of the complement of this set,
    i.e., $N$, has to be 0. Sets like these are called \emph{null
      sets}. A classical result says that \emph{null sets} do not need
    to be considered for the Lebesgue integral.
  \end{proof}
\end{lem}

For a well-behaved run, i.e., for every run that we need to consider
when calculating the expected value, we can calculate the payoff in
the following way.

\begin{lem}[Calculating the payoff of a well-behaved run]
  Let $\run$ be a well-behaved run of a unichain Markov chain.
  Denote by $\strategy: \mStates \to \Actions$ the only action
  available at a state. Then
  \[
  \payoff(\run) = \frac{\sum_{\mState
      \in \mStates} \mdpProb^*(\mState)
    \cost_1(\mState, \strategy(\mState))}
  {\lim_{l \to \infty} \frac{1}{l} + \sum_{\mState \in
      \mStates} \mdpProb^*(\mState) \cost_2(\mState, \strategy(\mState))}
  \]

  \begin{proof}
    By definition of $\payoff$ we have
    \[
    \payoff(\run) = \lim_{m \to\infty} \liminf_{l \to \infty}
    \frac{\sum_{i = l}^{m} \cost_1(\run_i)}{
         1 + \sum_{i =l}^{m} \cost_2(\run_i)}
    \]
    We now assume that the Markov chain consists of one maximal
    recurrence class. We can do this because every non-recurrent state
    will not influence $\payoff(\run)$, because $\run$ is well-behaved
    and because $\payoff$ is prefix independent. Hence
    \[
    \payoff(\run) = 
      \liminf_{l \to \infty} \frac{\sum_{i = 0}^{l} \cost_1(\run_i)}{1 + \sum_{i =
          0}^{l} \cost_2(\run_i)}
    \]
    We can calculate the sums in a different way: we take the
    sum over the states and count how often we visit one state, i.e.,
    \[
    \frac{\sum_{i = 0}^{l} \cost_1(\run_i)}{1 + \sum_{i =
          0}^{l} \cost_2(\run_i)} =
    \frac{
          \sum_{\mState \in \mStates} \cost_1(\mState, \strategy(\mState))
          \nVisits_{\mState}^l}{
          1 +
          \sum_{\mState \in \mStates} \cost_2(\mState, \strategy(\mState))
          \nVisits_{\mState}^l} =
    \frac{
          \sum_{\mState \in \mStates} \cost_1(\mState, \strategy(\mState))
          (\nVisits_{\mState}^l/l)}{
          1/l +
          \sum_{\mState \in \mStates} \cost_2(\mState, \strategy(\mState))
          (\nVisits_{\mState}^l/l)}
    \]

    Now we take $\lim$ instead of $\liminf$.  We will see later that
    the sequence converges for $l \to \infty$ and hence $\lim$ and
    $\liminf$ have the same value. Because both sides of the fraction
    are finite values we can safely draw the limit into the fraction,
    i.e.,
    \begin{align*}
    (\dagger) \lim_{l \to \infty} \left( \frac{
          \sum_{\mState \in \mStates} \cost_1(\mState, \strategy(\mState))
          (\nVisits_{\mState}^l/l)}{
          1/l +
          \sum_{\mState \in \mStates} \cost_2(\mState, \strategy(\mState))
          (\nVisits_{\mState}^l/l)} \right) = &
     \frac{\lim_{l \to \infty} 
          \left(
            \sum_{\mState \in \mStates} \cost_1(\mState, \strategy(\mState))
            (\nVisits_{\mState}^l/l)
          \right)
        }{
          \lim_{l \to \infty}
          \left(
           1/l +
           \sum_{\mState \in \mStates} \cost_2(\mState, \strategy(\mState))
           (\nVisits_{\mState}^l/l) \right)} \\ = &
    \frac{
          \sum_{\mState \in \mStates} \cost_1(\mState, \strategy(\mState))
          \lim_{l \to \infty} (\nVisits_{\mState}^l/l)}{
          \lim_{l \to \infty} (1/l) +
          \sum_{\mState \in \mStates} \cost_2(\mState, \strategy(\mState))
          \lim_{l \to \infty} (\nVisits_{\mState}^l/l)}
    \end{align*}
    
    Finally, by the definition of well-behaved runs we have $\lim_{l
      \to \infty} \frac{\nVisits_{\mState}^l}{l} =
    \prob^*(\mState)$. Hence
    \[
    \frac{
          \sum_{\mState \in \mStates} \cost_1(\mState, \strategy(\mState))
          \lim_{l \to \infty} (\nVisits_{\mState}^l/l)}{
          \lim_{l \to \infty} (1/l) +
          \sum_{\mState \in \mStates} \cost_2(\mState, \strategy(\mState))
          \lim_{l \to \infty} (\nVisits_{\mState}^l/l)} =
   \frac{
          \sum_{\mState \in \mStates} \cost_1(\mState, \strategy(\mState))
          \prob^*(\mState)}{
          \lim_{l \to \infty} (1/l) +
          \sum_{\mState \in \mStates} \cost_2(\mState, \strategy(\mState))
          \prob^*(\mState)}
    \]

    The limit diverges to $\infty$ if and only if the second costs
    are all equal to zero and at least one first cost is not. In this
    case the original definition of $\payoff$ diverges and hence
    $\payoff$ and the last expression are the same. Otherwise the
    last expression converges, hence $\dagger$ converges, ergo $\liminf$ and
    $\lim$ of this sequence are the same.
\end{proof}
\end{lem}

Note that the previous lemma implies that the value of a well-behaved
run is independent of the actual run. In other words, on the set of
well-behaved runs of a unichain Markov chain the payoff function
is constant\footnote{Note that the fact that any payoff function that
is prefix-independent is constant almost surely on each irreducible
Markov chain has already been proved by \cite{gimbert}}.
Ergo the expected value of such a Markov chain is equal
to the payoff of any of its well-behaved runs.

\begin{thm}[Expected payoff of a MDP and a strategy]
  Let $\MDP$ be a MDP  such that every pure and memoryless strategy induces an
  unichain MC. Let further
  $\prob^*$ denote the Cesaro limit of $\prob^n$ of the induced
  Markov chain. Then for every pure and memoryless strategy $\strategy$
  \[
    \eValue{_{\MDP}^{\strategy}}{\payoff} = \frac{
      \sum_{\mState \in \mStates} \cost_1(\mState, \strategy(\mState))
      \prob^*(\mState)}{
      \lim_{l \to \infty} (1/l) +
      \sum_{\mState \in \mStates} \cost_2(\mState, \strategy(\mState))
      \prob^*(\mState)}
  \]
  \begin{proof}
    This follows from the previous lemma and the fact that $\payoff$
    is constant on any well-behaved run.
  \end{proof}
\end{thm}

Note that this means that an expected value is $\infty$ if and only if
the second cost of every action in the recurrence class of the Markov
chain is 0 and there is at least one first cost that is not.

Using this lemma, we are now able to transform the MDP into a
fractional linear program. This is done in the same way as is done for
the expected average payoff case (cf.~\cite{puterman}).  We define
variables $x(\mState, \action)$ for every state $\mState \in \mStates$
and every available action $\action \in \ActionsOf(\mState)$. This
variable intuitively corresponds to the probability of being in state
$\mState$ and choosing action $\action$ at any time. Then we have, for
example $\prob^*(\mState) = \sum_{\action \in \ActionsOf(\mState)}
x(\mState, \action)$. We need to restrict this set of variables. First
of all, we always have to be in some state and choose some action,
i.e., the sum over all $x(\mState, \action)$ has to be one.  The
second set of restrictions ensures that we have a stationary
distribution, i.e., 
the sum of the probabilities of going out of (i.e., being in) a state
is equal to the sum of the probabilities of moving into this state.


\begin{defi}[Fractional Linear program for MDP]
  \label{def:lfp}
  Let $\MDP$ be an unichain MDP such that every Markov chain induced
  by any strategy contains at least one non-zero second cost. Then we
  define the following fractional linear program for it.

  \begin{equation}
    \label{eqn:lfp}
    \text{Minimize }    \frac{\sum_{\mState \in
      \mStates}\sum_{\action \in \ActionsOf(\mState)} 
    x(\mState, \action)\cost_1(\mState, \action)}{\sum_{\mState \in
      \mStates}\sum_{\action \in \ActionsOf(\mState)} x(\mState,
    \action) \cost_2(\mState, \action)}
  \end{equation}

  subject to

  \begin{eqnarray}
    \label{eqn:subjectTo1}
  & \sum_{\mState \in \mStates}\sum_{\action \in \ActionsOf(\mStates)}
  x(\mState, \action) = 1 \\
  \label{eqn:subjectTo2}
  & \sum_{\action \in \ActionsOf(\mState)} x(\mState, \action) = \sum_{\mState' \in
    \mStates} \sum_{\action \in \ActionsOf(\mState')} x(\mState',
  \action) \prob(\mState', \action)(\mState) & \forall \mState
  \in \mStates
\end{eqnarray}
\end{defi}

There is a correspondence between pure and memoryless
strategies and basic feasible solutions to the linear program\footnote{A
  feasible solution is one that fulfills the linear equations that
  every solution is subject to.}. That is, the linear program
always has a solution because every positional strategy corresponds to a
solution. See \cite{puterman} for a detailed
analysis of this in the expected average reward case.

Once we have calculated a solution of the linear program, we can
calculate the strategy as follows.

\begin{defi}[Strategy from solution of linear program]
  Let $x(\mState, \action)$ be the solutions to the linear
  program. Then we define the strategy as follows.

  \[ \strategy(\mState) = \begin{cases}
    \text{arbitrary} & \text{if } x(s, a) = 0 \text{ for every enabled
      action } a \\
    \action & \text{if } x(\mState, \action) > 0
    \end{cases} \]

  Note that this is well defined because for each state
  $\mState$ there is at most one action $\action$ such that
  $x(\mState, \action) > 0$ because of the bijection (modulo the action
  of transient states) between basic
  feasible solutions and strategies and because the optimal strategy
  is always pure and memoryless.
\end{defi}

\vspace{-3mm}
\subsection{From LFP to LP}

Since solvers to linear fractional programs are not common and there
are good free solvers to linear programs, we presented a method of
converting a linear fractional program to a sequence of linear
programs that calculate the solution. This algorithm is due to
\cite{IsbellMarlow}. Let $f(x)$ denote the value of \rEqn{eqn:lfp}
under variable assignment $x$.

\begin{algorithm}[H]
  \KwIn{feasible solution $x_0$, MDP $\MDP$}
  \KwOut{Variable assignment, optimal solution}

  $n \leftarrow 0$\\
  \Repeat{$f(x_{n-1}) = f(x_n)$}{
    $g \leftarrow f(x_n)$\\
    $n \leftarrow n + 1$\\
    Solve \[ \text{Minimize } \sum_{\mState \in \mStates} \sum_{\action \in
      \ActionsOf(\mState)} x_n(\mState, \action)c_{\mState}^1 - g
    \sum_{\mState \in \mStates}\sum_{\action \in \ActionsOf(\mState)}
    x_n(\mState, \action) c_{\mState}^2 \]
    subject to \rEqn{eqn:subjectTo1} and \rEqn{eqn:subjectTo2}.
  }
  \Return $x_n$, $f(x_n)$
\end{algorithm}

\vspace{-3mm}
\subsection{Preliminary Implementation}

We have developed a tool that can handle (finite) unichain MDPs with
ratio objectives based on the approach presented in this paper.  Our
tool is implemented in Haskell and uses the \emph{GNU Linear Programming
  Kit} to solve the resulting linear programs.

We made some initial experiments using the server-client example from
\rSec{sec:synthesis}.  In the case of two clients we have a MDP with
24 states and 288 edges. Building and solving this system takes
less than 100 milliseconds on a Laptop with an Intel Core 2 Duo P8600
clocked at 2.40 GHz.
The resulting machine behaves as follows:
If it receives only one request at the start, then it acknowledges
this request immediately.
Whenever Client~2, i.e., the complicated client, sends a request,
then it also receives the acknowledgment, with one exception:
When Client~1 has an outstanding request, i.e., if its qualitative
specification is in state $s_1$, and if Client~2 has no outstanding
request, then Client~1 receives the acknowledgment, 
The expected value is roughly $1.2 = 12/10$. This means that, out of
12 requests, 10 can be served, which means $83.3 \%$.

\vspace{-2mm}
\section{Conclusions and Future Work}
\vspace{-2mm}

We have presented a technique to automatically synthesize system that
satisfy a qualitative specification and optimize a quantitative
specification under a given environment model. Our technique can
handle qualitative specifications given by an automaton with a set of
safe states, and quantitative specifications defined by an automaton
with ratio objective.  

Currently, we are working on a better representation of the input
specifications. In particular, we are aiming for a symbolic
representation that would allow us to use a combined symbolic and
explicit approach, which has shown to be very effective for MDP with
long-run average objective~\cite{wimmer10}.  Furthermore, we are
extending the presented approach to qualitative specification describe
by arbitrary $\omega$-regular specifications.

\bibliographystyle{eptcs}
\bibliography{paper}

\end{document}